\documentclass[12pt]{article}
\usepackage{epsfig}
\usepackage{amsmath}
\usepackage{amssymb}
\usepackage{latexsym}
\usepackage{graphicx}
\newcommand{\be}{\begin{equation}}
\newcommand{\ee}{\end{equation}}
\newcommand{\bea}{\begin{eqnarray}}
\newcommand{\eea}{\end{eqnarray}}

\newcommand{\LL}{{\cal L}}

\newcommand{\TT}{{\cal T}}
\newcommand{\sectiono}[1]{\section{#1}\setcounter{equation}{0}}

\begin{document}
{}~ \hfill\vbox{\hbox{hep-th/0209197}\hbox{MIT-CTP-3310}}\break

\vskip 2.1cm

\centerline{\large \bf Stress Tensors in p-adic String Theory and
Truncated OSFT } \vspace*{4.0ex}

\centerline{\large \rm Haitang Yang }
\vspace*{4.0ex}
\centerline{\large \it Center for Theoretical Physics}

\centerline{\large \it Massachussetts Institute of Technology,}

\centerline{\large \it Cambridge, MA 02139, USA}
\centerline{E-mail: hyanga@mit.edu}

\vspace*{4.0ex} \centerline{\bf Abstract}
\medskip

We construct the stress tensors for the p-adic string model and
for the pure tachyonic sector of open string field theory by naive
metric covariantization of the action. Then we give the concrete
energy density of a lump solution of the p-adic model. In the
cubic open bosonic string field theory, we also give the energy
density of a lump solution and pressure evolution of a rolling
tachyon solution.

\baselineskip=16pt

\sectiono{Introduction}
Much work has been devoted to looking for solutions in string field
theory (SFT). Generally speaking, physicists are concerned with two
kinds of solutions with different properties. One kind of solutions are
the  time independent ones which represent the tachyon
vacuum or lower dimensional D-branes
\cite{9902105}-\cite{0008252}. Initiated by Sen \cite{0203211},
time dependent rolling tachyon solutions have recently attracted much
attention  \cite{0203265}-\cite{0209122}.
Studying rolling tachyon solutions can give us
information about how the tachyon approaches the tachyon vacuum. At
the same time, the p-adic model \cite{BFOW}, which
exhibits a lot of properties of string field theory, is also of interest.
In this model, the potential has a stable vacuum
and  a tachyon. Studying the dynamics of the tachyon
may suggest to us what happens in the same situation for the SFT.
Furthermore, one also has lump solutions in the p-adic theory
which are identified as lower dimensional D-branes \cite{0003278}.

In \cite{0207107}, Moeller and Zwiebach discussed how to construct
the stress tensor for the rolling tachyon solution in the p-adic
model. They obtained an unambiguous
expression for the energy through a generalized Noether procedure.
This analysis could not be extended to
the pressure calculation,  however, as there are ambiguities in that case.
Instead, they
included the metric in the action and used the definition of
stress tensor in general relativity to calculate the pressure.
Then they constructed the rolling tachyon solutions for both
the p-adic model and open string field theory (OSFT) in the form of
series expansions. After that, they calculated the pressure
evolution in the p-adic string case.

It is of interest to consider the stress tensor in the case when
the scalar field in the p-adic model depends on all the coordinates.
Especially, for a lump solution, what is the profile of the energy
distribution along the spatial coordinate? Is it the same as what
we expect intuitively? Furthermore, in OSFT,  it is important to
know if the profile of the energy density has the same properties
as that in p-adic string theory. Moeller and Zwiebach showed in
\cite{0207107} that the pressure of the rolling solution in p-adic
model does not vanish at large times. For the rolling solution in
OSFT, it is of interest to test if one gets vanishing pressure
asymptotically or not.

In this paper, we first give the stress tensor in a general form for
the p-adic model.
When our results are specialized to the time dependent solution
in p-adic model, they
reproduce the results in \cite{0207107}. A nontrivial lump
solution in p-adic model was given in \cite{BFOW}, \cite{0003278}.
We construct the energy density of this solution
and compare it with that of  the lump solution of ordinary
$\phi^3$ field theory. We find that these two energy densities
have similar spatial profiles.
Section $3$ is devoted to the case of the pure tachyon field in OSFT.
We again construct the stress
tensor in a general form.
The energy density of a  solitonic solution
\cite{0005036} is then constructed in subsection $3.1$. Finally we
calculate the pressure evolution of a rolling tachyon solution
\cite{0207107}.

\sectiono{p-adic String Theory Case}

In this section, we first construct the stress tensor of the
p-adic string theory by varying the metric. We will find that the
expression is exactly the same as the one obtained in
\cite{0207107} if we constrain scalar field to only depend on
time. We will also consider the case where the tachyon scalar only
depends one spatial coordinate. In that situation, one nontrivial
solitonic solution was already given \cite{BFOW}, \cite{0003278}.
We then calculate the energy density of that solution. The results
show that the total energy, integrated over all space, perfectly
agrees with the D24 brane tension as expected. The spatial profile
of this energy density looks very like the one of the solitonic
solution of ordinary $\phi^3$ field theory.
\subsection{Stress Tensor for p-adic model}
The p-adic string theory is defined by the action:
\begin{equation}
\label{pmodel} S = \int {d^d x\LL = \frac{1}{{g_p ^2 }}} \int {d^d
x} \left[ { - \frac{1}{2}\phi p^{ - \frac{1}{2}\Box } \phi  +
\frac{1}{{p + 1}}\phi ^{p + 1} } \right],\hspace{3mm}
\frac{1}{g_p^2}=\frac{1}{g^2}\frac{p^2}{p-1},
\end{equation}
where $\phi(x)$ is a scalar field, $p$ is a prime integer and $g$
is the open string coupling constant. Though the theory makes
sense even as $p\rightarrow 1$, in most cases, we will consider
$p\geq 2$ in this paper. In this action, there is an infinite
number of both time and spatial derivatives. One defines:
\begin{equation}
p^{ - \frac{1}{2}\Box }  \equiv \exp \left( { - \frac{1}{2}\ln
p\Box  } \right) = \sum\limits_{n = 0}^\infty {\left( { -
\frac{1}{2}\ln p} \right)} ^n \frac{1}{{n!}}\Box
^{n},
\end{equation}
and
\begin{equation}
\square = - \frac{{\partial ^2 }} {{\partial t^2 }} + \nabla
^2.
\end{equation}
Now we include the metric in the action \cite{0207107}:
\begin{eqnarray}
\label{gpmodel} S =S_1+S_2&=& \frac{1}{{g_p ^2 }}\int {d^d x} \sqrt { -
g} \left[ { - \frac{1}{2}\phi ^2  + \frac{1}{{p + 1}}\phi ^{p + 1}
} \right]\nonumber\\
&&- \frac{1}{{2g_p ^2 }}\sum\limits_{l = 1}^\infty {\left( { -
\frac{1}{2}\ln p} \right)^l \frac{1}{{l!}}} \int {d^d x} \sqrt { -
g} \phi \Box ^{l} \phi,
\end{eqnarray}
where we have split the
action into two parts: $S_1$ represents the potential and $S_2$
represents the kinetic term. After introduction of the metric
$\square$ becomes the covariant D'Alembertian.

\begin{eqnarray}
\label{covd} B_l\equiv\int d^dx \sqrt{-g} \phi\, \Box^l \,\phi &=&
\int d^dx\,\, \phi \,\, \partial_{\mu_1}
 \sqrt{-g}g^{\mu_1\nu_1}\partial_{\nu_1} \,\,
\,{1\over \sqrt{-g}} \partial_{\mu_2}
 \sqrt{-g}g^{\mu_2\nu_2}\partial_{\nu_2} \,\, \nonumber\\
&&  \cdots {1\over \sqrt{-g}} \partial_{\mu_l}
\sqrt{-g}g^{\mu_l\nu_l}\partial_{\nu_l} \, \phi \,.
\end{eqnarray}
The stress tensor is given by:
\begin{equation}
T_{\alpha \beta }  = \frac{2} {{\sqrt { - g} }}\frac{{\delta
S}} {{\delta g^{\alpha \beta } }}.
\end{equation}
The variation of the potential $S_1$ in ~(\ref{gpmodel})
contributes:
\begin{equation}
\label{vs1} \frac{2} {{\sqrt { - g} }}\frac{{\delta S_1}} {{\delta
g^{\alpha \beta } }}=- \frac{1} {{g_p ^2 }}\left( { - \frac{1}
{2}\phi ^2  + \frac{1} {{p + 1}}\phi ^{p + 1} } \right)g_{\alpha
\beta },
\end{equation}
where we have set the metric to be flat with signature
$(-,+,+\cdots +)$ after the variation and we will use the same
convention in the rest of this paper. As for the variation of the
kinetic term $S_2$ in ~(\ref{gpmodel}), from ~(\ref{covd}), we
need to vary both factors of $\sqrt{-g}$ and $g^{\mu_i\nu_i}$ with
respect to $g^{\alpha\beta}$. First consider varying factors of
$\sqrt{-g}$ in ~(\ref{covd}) with respect to $g^{\alpha\beta}$:
\begin{eqnarray}
\frac{\delta B_l}{\delta \sqrt{-g}} \frac{\delta\sqrt{-g}}{\delta
g^{\alpha\beta}}&=& g^{\mu _1 \nu _1 } g^{\mu _2 \nu _2 }  \cdots
g^{\mu _l \nu _l } ( \phi _{\mu _1 } \phi _{\nu _1 \mu _2 \nu _2
\cdots \mu _l \nu _l }  + \phi _{\mu _1 \nu _1 } \phi _{\mu _2 \nu
_2 \cdots \mu _l \nu _l }  +  \cdots  \nonumber\\
&&\cdots+ \phi _{\mu _1 \nu _1 \mu _2 \nu _2 \cdots \mu _l } \phi
_{\nu _l }  )g_{\alpha \beta },
\end{eqnarray}
with the definition:
\[
\phi _{\mu _1 \nu _1 \mu _2 \nu _2  \cdots \mu _l \nu _l }  \equiv
\partial_{\mu _1 }\partial_{\nu
_1 }\partial_{ \mu _2 }\partial_{\nu _2 }\cdots \partial_{\mu _l
}\partial_{\nu _l }\phi (x).
\]
The variation of the factors of $g^{\mu_i\nu_i}$ in ~(\ref{covd})
with respect to $g^{\alpha\beta}$ contributes:
\begin{eqnarray}
\frac{\delta B_l}{\delta g^{\mu_i\nu_i}}\frac{\delta
g^{\mu_i\nu_i}} {\delta g^{\alpha\beta}}&=&- 2g^{\mu _1 \nu _1 }
g^{\mu _2 \nu _2 } \cdots g^{\mu _{l - 1} \nu _{l - 1} } \Big(\phi
_\alpha  \phi _{\beta \mu _1 \nu _1 \mu _2 \nu _2 \cdots \mu _{l -
1} \nu _{l - 1} } \nonumber \\ &&+ \phi _{\alpha \mu _1 \nu _1 }
\phi _{\beta \mu _2 \nu _2 \cdots \mu _{l - 1} \nu _{l - 1} } +
\cdots  + \phi _{\alpha \mu _1 \nu _1 \mu _2 \nu _2 \cdots \mu _l
} \phi _\beta \Big).
\end{eqnarray}
So, we can calculate $\delta S_2$. Finally, the stress tensor is:
\begin{eqnarray}
\label{padicT}T_{\alpha\beta}&=&- \frac{1} {{g_p ^2 }}\left(
{ - \frac{1} {2}\phi ^2  + \frac{1} {{p + 1}}\phi ^{p + 1} }
\right)g_{\alpha
\beta }\nonumber\\
&&-\frac{1}{{2g_p ^2 }}\sum\limits_{l = 1}^\infty  {\left( { -
\frac{1}{2}\ln p} \right)^l } \frac{1}{{l!}}\bigg\{g^{\mu _1 \nu
_1 } g^{\mu _2 \nu _2 }  \cdots g^{\mu _l \nu _l }\Big( \phi _{\mu _1
} \phi _{\nu _1 \mu _2 \nu _2  \cdots \mu _l \nu _l }\nonumber\\
&& + \phi_{\mu _1 \nu _1 } \phi _{\mu _2 \nu _2
\cdots \mu _l \nu _l } +
\cdots+ \phi _{\mu _1 \nu _1 \mu _2 \nu _2 \cdots \mu _l } \phi
_{\nu _l } \Big )g_{\alpha \beta }\nonumber\\
&& - 2g^{\mu _1 \nu _1 } g^{\mu _2 \nu
_2 }  \cdots g^{\mu _{l - 1} \nu _{l - 1}}\Big(\phi
_\alpha \phi _{\beta \mu _1 \nu _1 \mu _2 \nu _2  \cdots \mu _{l -1}
\nu _{l - 1} }\nonumber\\
&& + \phi _{\alpha \mu _1 \nu _1 } \phi _{\beta \mu _2 \nu _2
\cdots \mu _{l - 1} \nu _{l - 1} }+ \cdots  + \phi _{\alpha \mu _1
\nu _1 \mu _2 \nu _2  \cdots \mu _l } \phi _\beta \Big
)\bigg\}.
\end{eqnarray}
If $\phi(x)$ is only time dependent, in (\ref{padicT}), each
$g^{\mu_i\nu_i}$ contributes one `$-$' sign and the second term in
the sum survives only for the component $T_{00}$. This gives the
same results as in ~\cite{0207107}.

One can also use the following identity \footnote{I thank M. Schnabl for
suggesting the use of this identity.}
\[
\delta e^A=\int_0^1 dt\, e^{tA}(\delta A) e^{(1-t)A}
\]
to get an alternative ``closed'' form of the stress tensor, compared
with the series expression (\ref{padicT}):
\begin{eqnarray}
\label{padicT2} T_{\alpha \beta } &=&
\frac{{g_{\alpha \beta } }} {{2g_p ^2 }}\bigg\{\phi e^{ - k\square
} \phi  - \frac{2} {{p + 1}}\phi ^{p + 1}  + k\int\limits_0^1 {dt}
(e^{ - kt\square } \phi )(\square e^{ - k(1 - t)\square } \phi
)\nonumber\\
&&+ k\int\limits_0^1 {dt} (\partial _\mu  e^{ - kt\square } \phi )
(\partial ^\mu  e^{ - k(1 - t)\square } \phi )\bigg\}\nonumber\\
&&- \frac{k} {{g_p ^2 }}\int\limits_0^1 {dt} (\partial _\alpha e^{
- kt\square } \phi )(\partial _\beta  e^{ - k(1 - t)\square }
\phi),
\end{eqnarray}
where $k\equiv \frac{1}{2}\ln p$.

In the case that $\phi(x)$ only depends on one spatial coordinate,
say $x\equiv x^{25}$, the last term in the right hand side of
(\ref{padicT2}) vanishes for all the components except for
$T_{25,25}$. The energy density is

\begin{eqnarray}
\label{peng} E(x)&=&T^0_0=
\frac{1} {{2g_p ^2 }}\bigg\{\phi e^{ - k\partial^2 } \phi  -
\frac{2} {{p + 1}}\phi ^{p + 1}  + k\int\limits_0^1 {dt} (e^{ -
kt\partial^2 } \phi )(\partial^2 e^{ - k(1 - t)\partial^2 } \phi
)\nonumber\\
&&+ k\int\limits_0^1 {dt} (\partial   e^{ - kt\partial^2 } \phi )
(\partial   e^{ - k(1 - t)\partial^2 } \phi )\bigg\},
\end{eqnarray}
where $\partial^2\equiv\frac{\partial^2}{\partial x^2}$.

\subsection{Energy of The Lump Solution}

\begin{figure}
\centerline{\hbox{\epsfig{figure=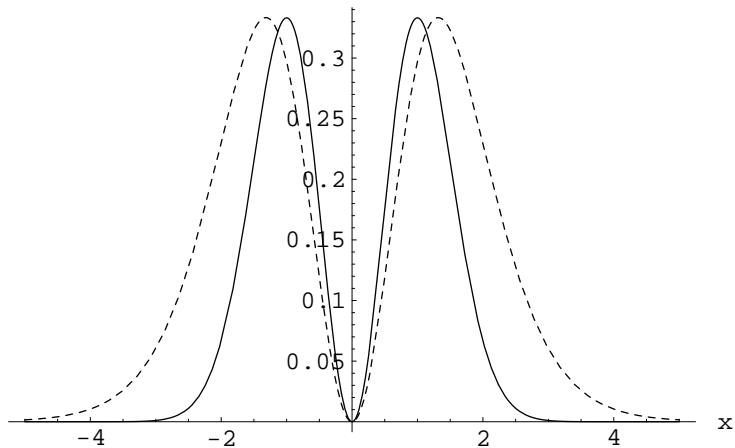, height=6cm}}}
\caption{The energy distribution of the lump solution
(\ref{plump}) of the p-adic model for $p=2$ (solid line, $g_p^2
E(x)$ versus $x$) and that (\ref{phi3lump}) of ordinary $\phi^3$
field theory (dashed line, $g_0^2E(x)$ versus $x$).}
\label{padiclumpE}
\end{figure}

There are  some previously known solutions for the p-adic model
\cite{BFOW}, \cite{0003278}. One of them is the lump solution:
\begin{equation}
\label{plump} \phi (x) = p^{\frac{1}{{2(p - 1)}}} \exp \left( { -
\frac{1}{2}\frac{{p - 1}}{{p\ln p}}x^2 }\right).
\end{equation}
This solution is interpreted as a D24-brane, where $x$ is the
coordinate transverse to the brane. This solution can be
generalized to lower dimensional branes \cite{0003278}. The
D-brane tension of this solution is:
\begin{eqnarray}
\label{tension} \TT_{24}&=&-\int dx \LL (\phi(x))
=-\int dx\frac{1}{2g_p^2}\frac{1-p}{1+p}\phi^{(p+1)}(x)\nonumber\\
&=&\frac{1}{g_p^2}\frac{p-1}{2(p+1)}p^{\frac{p}{p-1}}\sqrt\frac{2\pi\ln
p}{p^2-1}.
\end{eqnarray}
Using the identity
\[
\exp\left(-a\frac{d^2}{dx^2}\right)\exp(-bx^2)=\frac{1}
{\sqrt{1-4ab}}\exp\left(-\frac{bx^2}{1-4ab}\right),
\]
from (\ref{peng}), we can write down the energy density:
\begin{equation}
\label{pengs} E(x)=\frac{p-1}{p+1}\sqrt{\frac{2\pi}{(p^2-1)\ln
p}}\, p^{\frac{p}{p-1}}|x| \,{\rm
Erf}\left[\frac{p-1}{p+1}\sqrt{\frac{p^2-1}{2p\ln p}}\,|x|\right]
e^{-\frac{2(p-1)x^2}{(p+1)\ln p}},
\end{equation}
where ${\rm Erf}[x]\equiv\frac{2}{\sqrt\pi}\int_0^x
dt\,\exp(-t^2)$ is the error function. In Figure \ref{padiclumpE},
we plot this energy density (the solid line) for $p=2$. At $x=0$
and $x\rightarrow\pm\infty$, this energy density vanishes. By
solving $\frac{d}{dx}E(x)=0$ numerically as $p=2$, one can see the
energy reaches its maxima at $x\approx\pm0.9997$. From
(\ref{pmodel}), the potential is
\[
\frac{1}{2}\phi^2-\frac{1}{p+1}\phi^{p+1},
\]
so, the D-brane vacuum is at $\phi=1$. Moreover, from
(\ref{plump}), one gets $\phi=1$ at $x=\pm\sqrt 2 \ln2\approx \pm
0.9803$, which are close to the locations where the energy gets
its maxima.

The lump solution (\ref{plump}) we are considering here, as we
mentioned at the beginning of this subsection, is interpreted
as a D-24 brane sharply localized on the
hyperplane $x=0$. Therefore,  intuitively one may expect
the energy to be sharply localized around $x=0$. But from figure
\ref{padiclumpE}, one can see that the energy is somewhat localised
around $x\approx\pm 0.9997$ and reaches a local minimum at $x=0$.

The total energy is:
\begin{equation}
\label{PTE} \int\limits_{-\infty}^{\infty} dx\,
E(x)=\frac{1}{g_p^2}\frac{p-1}{2(p+1)}p^{\frac{p}{p-1}}\sqrt\frac{2\pi\ln
p}{p^2-1}
\end{equation}
which is exactly the same as (\ref{tension}). In the limit
$p\rightarrow 1$, $E(x)$ becomes:
\begin{equation}
\label{limite}\mathop {\lim }\limits_{p \to 1} E(x) =
\frac{1}{{2g^2 }}x^2 \exp (1 - x^2 ).
\end{equation}
On the other hand, from (\ref{pmodel}), as $p\rightarrow 1$, the
action becomes:
\[
S=\frac{1}{2g^2}\int d^d x \bigg( \frac{1}{2}\phi\square\phi
-\frac{1}{2}\phi^2+\phi^2 \ln\phi\bigg).
\]
This action has a lump solution:
\[
\phi(x)=\exp\Big(\frac{1}{2}(1-x^2)\Big),
\]
whose energy density is exactly the same as (\ref{limite}).

This energy density looks very similar to that of the ordinary
$\phi^3$ field theory with coupling constant $g_0$ and unit mass
\cite{0008227}:
\[
S=\frac{1}{g_0^2}\int d^d x\bigg\{\frac{1}{2}(\partial\phi)^2-\frac{1}{2}
\phi^2+\frac{1}{3}\phi^3\bigg\},
\]
which has the lump solution:
\begin{equation}
\label{phi3lump}\phi (x)=\frac{3}{2}(1-{\rm tanh}^2\frac{x}{2}),
\end{equation}
with energy density
\begin{equation}
E(x)=\frac{1}{g_0^2}\frac{9}{4}{\rm sech}^4\frac{x}{2} {\rm
tanh}^2\frac{x}{2},
\end{equation}
which is plotted in Figure 1 (dashed line).

\sectiono{The Pure Tachyon Field of String Field Theory Case}
When we expand the string field in the Hilbert space of the first
quantized string theory, we can read off the action of the pure
tachyonic cubic string field theory. As in the last section, we
include the metric in the action and convert all the ordinary
derivatives to covariant ones. Variations of the metric again give
the stress tensor. Then we calculate the energy density of
the lump solution given in \cite{0005036} and the pressure
of the rolling tachyon solution given in \cite{0207107}.

\subsection{Stress Tensor for the Tachyon field in SFT}
Firstly, we write down the pure tachyonic action of the cubic SFT.
From Sen's conjecture \cite{9902105}, we should add the D-brane
tension into the SFT action to cancel the negative energy due to
the tachyon.   We know that after adding the D-brane tension term
to the potential of the cubic SFT, the local minimum of the new
potential vanishes \cite{9912249}. In the same spirit, here we
should add a term $\frac{1}{6} K^{-6}$ to the potential to set the
local minimum of the potential to zero.
\begin{equation}
\label{sftaction} S = \frac{1}{{g_0 ^2 }}\int {d^d x} \left(
{\frac{1}{2}\phi ^2 - \frac{1}{2}(\partial \phi )^2  -
\frac{1}{3}K^3 \widetilde\phi ^3 -\frac{1}{6} K^{-6}} \right),
\end{equation}
where
\begin{equation}
\label{dphitild} \widetilde\phi  = \exp \left( {\ln K\square }
\right)\phi (x) = K^\square  \phi (x).
\end{equation}
$g_0$ is the open bosonic string coupling constant and $K = 3\sqrt
3 /4$. $\square$ is defined as in the last section. The equation
of motion from this action is:
 \[
 K^{ - 2\square } (1 + \square )\widetilde\phi  = K^3
\widetilde\phi ^2.
\]
In order to separate the term without
derivatives from $\widetilde\phi (x)$, we define:
\begin{eqnarray}
\label{dpsi}\psi (x) &=& \widetilde\phi (x) - \phi (x) =
\sum\limits_{l = 1}^\infty {\frac{{\left( {\ln K} \right)^l }}
{{l!}}} \square ^l \phi (x)\nonumber\\
&=& \sum\limits_{l = 1}^\infty  {\frac{{\left( {\ln K} \right)^l
}} {{l!}}} \frac{1} {{\sqrt { - g} }}\partial _{\mu _1 } \sqrt { -
g} g^{\mu _1 \nu _1 } \partial _{\nu _1 } \frac{1} {{\sqrt { - g}
}}\partial _{\mu _2 } \sqrt { - g} g^{\mu _2 \nu _2 } \partial
_{\nu _2 }  \cdots\nonumber\\
&&\cdots \frac{1} {{\sqrt { - g} }}\partial _{\mu _l } \sqrt { -
g} g^{\mu _l \nu _l } \partial _{\nu _l } \phi (x),
\end{eqnarray}
where in the last step, we have written the expression in the
covariant form. For an arbitrary differentiable function $f(x)$,
\begin{equation}
\label{general}\int {d^d xf(x)} \frac{\delta\psi (x)}{\delta
g^{\alpha \beta } }= \frac{1} {2}f\psi g_{\alpha \beta
}+A_{\alpha\beta}(f)
\end{equation}
where
\begin{eqnarray}
\label{defineA} A_{\alpha\beta}(f)&=& \frac{1}
{2}g_{\alpha \beta } \sum\limits_{l = 1}^\infty  {\frac{{\left(
{\ln K} \right)^l }} {{l!}}} g^{\mu _1 \nu _1 }  \cdots g^{\mu _l
\nu _l }
\nonumber\\
&&\cdot\Big( {f_{\mu _1 } } \phi _{\nu _1 \mu _2 \nu _2  \cdots
\mu _l \nu _l }
 + f_{\mu _1 \nu _1 } \phi _{\mu _2 \nu _2  \cdots \mu _l \nu _l
}  +  \cdots + {f_{\mu _1 \nu _1  \cdots \mu _l } \phi _{\nu
_l } } \Big )\nonumber\\
&& - \sum\limits_{l = 1}^\infty  {\frac{{\left( {\ln K} \right)^l
}} {{l!}}} g^{\mu _1 \nu _1 }  \cdots g^{\mu _{l - 1} \nu _{l - 1}
} \Big( {f_\alpha  } \phi _{\beta \mu _1 \nu _1  \cdots
\mu _{l - 1} \nu _{l - 1} }  +\nonumber\\&&
 f_{\alpha \mu _1 \nu_1 }
\phi _{\beta \mu _2 \nu _2  \cdots \mu _{l - 1} \nu _{l - 1} } +
\cdots  +  {f_{\alpha \mu _1 \nu _1  \cdots \mu _{l - 1} \nu _{l -
1} } \phi _\beta  } \Big).
\end{eqnarray}
Again, we set the metric to be flat with signature $(-1,1,1\cdots
1)$ after the variation. Replace $\widetilde\phi$ by $\phi+\psi$
in (\ref{sftaction}), expanding and coupling to the metric:
\begin{eqnarray}
\label{SFTag}S &=& \frac{1} {{g_0 ^2 }}\int {d^d x} \sqrt {
- g} \left\{ {\left( {\frac{1} {2}\phi ^2  - \frac{1}
{2}g^{\mu\nu}\partial_\mu \phi\partial_\nu\phi  - \frac{1}
{3}K^3 \phi ^3-\frac{1}{6} K^{-6} } \right)} \right.\nonumber\\
&& \left. { - K^3 \left( {\phi ^2 \psi  + \phi \psi ^2  + \frac{1}
{3}\psi ^3 } \right)} \right\}.
\end{eqnarray}
Varying the first term in the last right hand side of
(\ref{SFTag}) with respect to $\delta g^{\alpha\beta}$ gives
\begin{eqnarray}
\label{svs1}&&\frac{1} {{g_0 ^2
}}\delta_{g^{\alpha\beta}}\int {d^d x} \sqrt { - g} {\left(
{\frac{1} {2}\phi ^2  - \frac{1} {2}g^{\mu\nu}\partial_\mu
\phi\partial_\nu\phi  - \frac{1} {3}K^3
\phi ^3-\frac{1}{6} K^{-6} } \right)} \nonumber\\
 &=&- \frac{{g_{_{\alpha \beta } } }}
{{2g_0 ^2 }}\left( {\frac{1} {2}\phi ^2  - \frac{1} {2}(\partial
\phi )^2  - \frac{1} {3}K^3 \phi ^3 }-\frac{1}{6} K^{-6} \right) -
\frac{1} {2g_0^2}\partial _\alpha  \phi \partial _\beta \phi\nonumber\\
&\equiv&-\frac{1}{g_0^2}C_{\alpha\beta},
\end{eqnarray}
where we have defined $C_{\alpha\beta}$ to simplify our notation.
As for the second term in the last right hand side of
(\ref{SFTag}), note
\begin{eqnarray*}
&&-K^3\delta _{g^{\alpha \beta } } \int {d^d x} \sqrt { - g} \left(\phi ^2 \psi
+\phi\psi^2+\frac{1}{3}\psi^3 \right)\\
&&= \frac{1} {2}K^3 \Big(\phi ^2 \psi+\phi\psi^2+\frac{1}{3}\psi^3\Big)
g_{_{\alpha \beta } }  - K^3 \int {d^d x} \sqrt { - g}\widetilde\phi^2
\delta _{g^{\alpha \beta } } \psi.
\end{eqnarray*}
So, from (\ref{general}) and (\ref{defineA}) the variation of the
second term in the last step of (\ref{SFTag}) contributes:
\begin{eqnarray}
\label{svs2} && \left( \frac{{ - K^3 }}{g_0^2} \right)\delta
_{g^{\alpha \beta } }\int {d^d x} \sqrt { - g} \left( {\phi ^2
\psi  + \phi \psi ^2  + \frac{1} {3}\psi ^3 } \right)\nonumber\\
&&=\frac{-K^3}{g_0^2}\left\{ A_{\alpha \beta }(\widetilde\phi ^2)
+ \frac{1} {2}\left( {\phi \psi ^2  + \frac{2} {3}\psi ^3 }
\right)g_{\alpha \beta }\right\},
\end{eqnarray}
Finally, from (\ref{defineA}), (\ref{svs1}) and (\ref{svs2}), the
stress tensor is:
\begin{equation}
\label{SFTstress} T_{\alpha \beta } = \frac{2} {{\sqrt { -
g}}}\frac{{\delta S}} {{\delta g^{\alpha \beta } }} =-
\frac{2K^{-3}} {{g_0 ^2 }}A_{\alpha \beta
}(\widetilde\phi^2)-\frac{2}{g_0^2}C_{\alpha\beta} .
\end{equation}
In the case that $\phi (x)$ only depends on one spatial
coordinate, say $x^{25}$, from (\ref{defineA}),
\begin{equation}
\label{lumpA} A_{\alpha \beta }(\widetilde\phi^2)=\sum\limits_{l =
1}^\infty {\frac{{\left( {\ln K} \right)^l }} {{l!}}} \left\{
{\frac{1} {2}g_{\alpha \beta } \sum\limits_{m = 1}^{2l - 1}
{\widetilde\phi ^2 _m } \phi _{2l - m}  - \delta _{\alpha,25}
\delta _{\beta,25} \sum\limits_{m = 1}^l {\widetilde\phi ^2 _{2m -
1} \phi _{2l - 2m + 1} } } \right\}.
\end{equation}
Plug it into (\ref{SFTstress}), we obtain the stress tensor for
lump solutions. Similarly, if $\phi (x)$ only depends on time, we
can write:
\begin{equation}
\label{rollA} A_{\alpha \beta }(\widetilde\phi^2)  =  - K^3
\sum\limits_{l = 1}^\infty {\frac{{\left( { - \ln K} \right)^l }}
{{l!}}} \left\{ {\frac{1} {2}g_{\alpha \beta } \sum\limits_{m =
1}^{2l - 1} {\widetilde\phi ^2 _m } \phi _{2l - m}  + \delta
_{\alpha,0} \delta _{\beta,0} \sum\limits_{m = 1}^l
{\widetilde\phi ^2 _{2m - 1} \phi _{2l - 2m + 1} } }
\right\}.
\end{equation}
Plug it into (\ref{SFTstress}), we obtain the stress tensor for
rolling solutions

\subsection{Energy distribution of the SFT lump solution}
In \cite{0005036}, a lump solution of OSFT has been given in the
form of an expansion in terms of cosines. We are only concerned
with the pure tachyonic mode here, so drop the higher modes:
\begin{equation}
\label{SFTlumpsolution} \phi (x) = t_0  + t_1 \cos \left(
{\frac{x} {R}} \right) + t_2 \cos \left( {\frac{{2x}} {R}} \right)
+  \cdots,
\end{equation}
where $R$ is the radius of the circle on which the coordinate $x$
is compactified. We can calculate the energy distribution of this
solution, from (\ref{dphitild}), (\ref{dpsi}), (\ref{SFTstress})
and (\ref{lumpA}):
\[
\widetilde\phi (x) = K^{\partial _x ^2 } \phi (x) = t_0  + t_1 K^{
- \frac{1} {{R^2 }}} \cos \left( {\frac{x} {R}} \right) + t_2 K^{
- \frac{4} {{R^2 }}} \cos \left( {\frac{{2x}} {R}} \right) +
\cdots,
\]
\[
\psi (x) = \widetilde\phi (x) - \phi (x) = t_1 \left( {K^{ -
\frac{1} {{R^2 }}}  - 1} \right)\cos \left( {\frac{x} {R}} \right)
+ t_2 \left( {K^{ - \frac{4} {{R^2 }}}  - 1} \right)\cos \left(
{\frac{{2x}} {R}} \right) +  \cdots,
\]

\begin{eqnarray}
\label{SFTE}E(x) &=& T^0 _0  =  - T_{00}\nonumber\\
&=&  - \frac{1} {{g_0
^2 }}\left( {\frac{1} {2}\phi ^2  - \frac{1} {3}K^3 \phi ^3  -
\frac{1} {2}\left( {\partial _x \phi } \right)^2  - \frac{1}
{6}K^{ - 6}+K^3 \phi \psi ^2  + \frac{2}{3}K^3 \psi ^3}\right)\nonumber\\
&& - \frac{{K^3 }} {{g_0 ^2 }}\sum\limits_{l = 1}^\infty
{\frac{{\left( {\ln K} \right)^l }} {{l!}}} \sum\limits_{m =
1}^{2l - 1} {\left( {\widetilde\phi ^2 } \right)_m } \phi _{2l -
m}.
\end{eqnarray}
In $R=\sqrt 3$ case, using the method introduced in
\cite{0005036}, one can obtain:
\[
t_0=0.216046,\hspace{5mm}t_1=-0.343268,\hspace{5mm}t_2=-0.0978441,
\]
when we plug these values into (\ref{SFTE}), we find:
\begin{eqnarray*}
E(x)&=&\frac{1}{g_0^2}\Big(0.0206937
+0.0242345 \cos \frac{x}{R}\\
&&-0.00780954 \cos \frac{2x}{R}  -0.0204855 \cos
\frac{3x}{R}\\
&&-0.0111187\cos \frac{4x}{R}-0.00218278\cos
\frac{5x}{R}-0.000177055\cos \frac{6x}{R}\Big).
\end{eqnarray*}
This lump solution has the interpretation of D24 brane, the
tension is:
\[
{\cal T}_{24}=\int\limits_{-\pi R}^{\pi R} dx\, E(x)\simeq 0.225206 \frac{1}{g_0^2}.\]
On the other hand, $\phi=0$ is supposed to represent the D25 brane. We have ${\cal
T}_{25}=-V(\phi=0)=\frac{1}{6}\frac{K^{-6}}{g_0^2}\simeq
0.0346831\frac{1}{g_0^2}$. Therefore,
\[\frac{1}{2\pi}\frac{{\cal T}_{24}}{{\cal T}_{25}}\simeq 1.03343
\]
a ratio that is unity in string theory.

Figure 2 shows the energy density $E(x)$. As the lump solutions in
the p-adic string theory, the energy density is not localised
around the hyperplane $x=0$. Instead, $E(x=0)$ is a local minimum.
A difference from the p-adic model is that $E(0)$ does not vanish
here.

\begin{figure}
\centerline{\hbox{\epsfig{figure=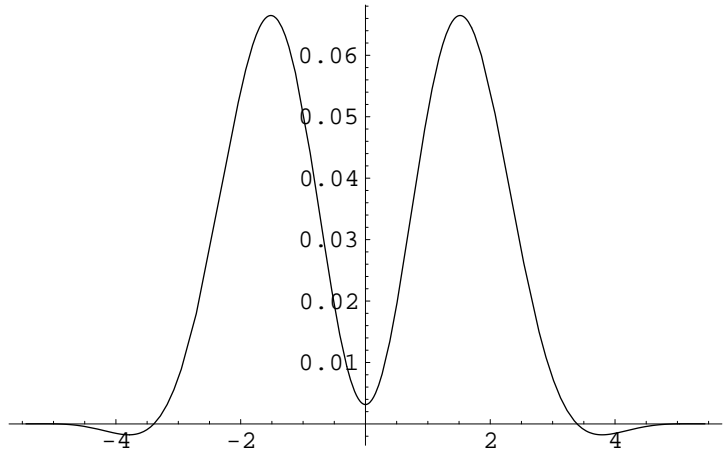, height=6cm}}}
\caption{The Energy density of the pure tachyonic lump solution
$\phi(x)=t_0+t_1{\rm cos}\frac{x}{R}+t_2{\rm cos}\frac{2x}{R}$ of
OSFT theory with $R=\sqrt 3$. the plot is  $g_0^2 E(x)$ versus $x$
.} \label{0207SFTlumpE}
\end{figure}

\subsection{Pressure evolution of the SFT rolling tachyon solution}

\begin{figure}
\centerline{\hbox{\epsfig{figure=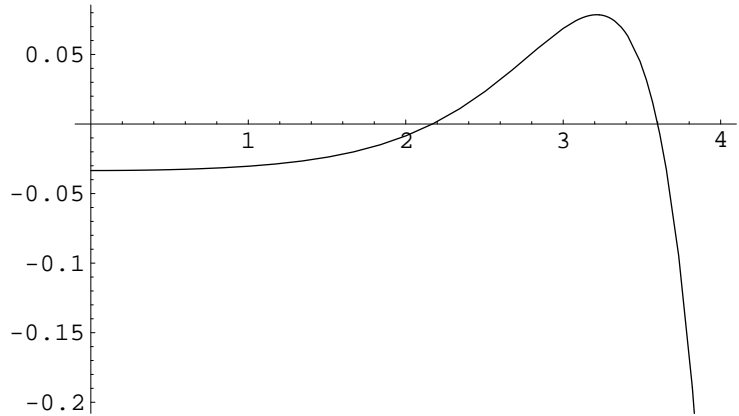, height=6cm}}}
\caption{The pressure evolvement of the rolling tachyon  solution
$\phi(t)=t_0+t_1\cosh t+t_2\cosh 2t$ of OSFT theory. $g_0^2 p(t)$
versus $t$ . As t becomes larger, $p(t)$ oscillates rapidly.}
\label{0207SFTrollP}
\end{figure}

In \cite{0207107}, a rolling tachyon solution of OSFT is expressed
as a series expansion in cosh($nt$):
\[\phi(t)=t_0+t_1{\rm cosh}t+t_2{\rm cosh}2t+\cdots.\]
From (\ref{dphitild}), (\ref{dpsi}), (\ref{SFTstress}) and (\ref{rollA}):
\[\widetilde\phi (t) = K^{ - \partial _t ^2 } \phi (t) = t_0  + t_1
K^{ - 1} \cosh t  + t_2
K^{ - 4} \cosh 2t  +
\cdots,\]
\[\psi (t) = \widetilde\phi (t) - \phi (t) = t_1 \left( K^{ -
1}  - 1 \right)\cosh t
+ t_2 \left (K^{ - 4}  - 1\right)\cosh
2t +  \cdots,\]

\begin{eqnarray}
p(t) &=&  - T_{11}\nonumber\\
&=& \frac{1} {{g_0 ^2 }}\left( {\frac{1}
{2}\phi ^2  - \frac{1} {3}K^3 \phi ^3  + \frac{1} {2}\left(
{\partial _t \phi } \right)^2  - \frac{1} {6}K^{ - 6}  + K^3 \phi
\psi ^2  + \frac{2} {3}K^3 \psi ^3 } \right)\nonumber\\
&& +\frac{{K^3 }} {{g_0 ^2 }}\sum\limits_{l = 1}^\infty
{\frac{{\left( { - \ln K} \right)^l }} {{l!}}} \sum\limits_{m =
1}^{2l - 1} {\left( {\widetilde\phi ^2 } \right)_m } \phi _{2l -
m}.
\end{eqnarray}
From  section 7 in \cite{0207107},
\[t_0=0.00162997,\hspace{5mm}t_1=0.05,\hspace{5mm}t_2=-0.000189714,\]
and therefore,

\begin{eqnarray*}p(t)&=&\frac{1}{g_0^2}\Big(-0.0346844+0.0000416895\cosh t
+0.00124462\cosh 2t\\
&&-0.0000416042\cosh 3t
+2.59666\times 10^{-7}\cosh 4t\\
&&-3.97466\times
10^{-10}\cosh 5t +2.09045\times 10^{-13}
\cosh 6t\Big).
\end{eqnarray*}
Figure \ref{0207SFTrollP} shows the pressure evolution. It has the same property as
the pressure in p-adic theory (Figure 10 in \cite{0207107}).The pressure
starts from negative value at time $t=0$ to force
the tachyon roll to the vacuum. But instead of decreasing to zero as $t\rightarrow
\infty$, it oscillates without bound at large times. So, this solution does not seem to
represent tachyon matter.

\sectiono{Conclusion}

By introducing the metric, we have obtained general expressions for
the stress tensors both for the p-adic model and for the pure tachyonic sector of
open bosonic string field theory \cite{9902105}, \cite{9912249}, \cite{0002237},
\cite{0005036}.

Furthermore, we considered some available solutions and wrote down
the corresponding energy densities for space dependent ones and
pressure evolutions for time dependent ones. In conformal field theory,
D-branes are boundary conditions and one could expect the energy to be sharply localized
at the D-brane position. It was not clear whether or not the lumps of the padic
string theory would have this property. Our results show that they do not.
The energy density vanishes at
$x=0,\pm\infty$. It has two maxima. These two maxima are
symmetrically localized with respect to $x=0$. In the pure
tachyonic sector of OSFT, the energy density for the lump solution
reaches a local minimum at $x=0$.
For the rolling tachyon solution, the pressure
oscillates with growing amplitude instead of asymptotically
vanishing. Therefore, as in the p-adic model, the rolling solution
we considered in this paper does not seem to represent tachyon
matter.

There are two shortcomings of the calculations in OSFT. The first
is not including the massive fields. The second is that the
coupling of open strings to the metric could have additional terms
that vanish in the flat space limit but contribute to the stress
tensor. Such phenomena happens in noncommutative field theory
\cite{0012218}. Open-closed string field theory \cite{9705241} might be needed to
calculate the stress tensor with complete confidence. I thank M. Schnabl for
bringing this point to my attention.

\bigskip
\noindent {\bf Acknowledgements.} The author is especially
grateful to B. Zwiebach and M. Schnabl for their critical help in
finishing this work. Also I thank Jessie Shelton and Nikhil Mittal
for useful discussions. This work was supported by DOE contract
\#DE-FC02-94ER40818.

\end{document}